\documentclass[accepted]{bmaw2022} 

\usepackage[american]{babel}

\usepackage{natbib} 
\usepackage{mathtools} 
\usepackage{booktabs} 
\usepackage{color}
\usepackage{verbatim}
\usepackage{etoolbox}
\usepackage{listings}

\title{Hierarchical Bayesian Regression for Multi-Location Sales Transaction Forecasting}
\author[1]{\href{mailto:<john-mark.agosta@microsoft.com>?Subject=Your UAI 2022 paper}{John~Mark~Agosta}{}}
\author[1]{{}{Mario~Inchiosa}{}}
\affil[1]{%
    Azure Data\\
    Microsoft\\
    Mountain View, CA, USA
}

\begin{document}
\maketitle

\begin{abstract}
    The features in many prediction models naturally take the form of a hierarchy. The lower levels represent individuals or events. These units group naturally into locations and intervals or other aggregates, often at multiple levels. Levels of groupings may intersect and join, much as relational database tables do.  Besides representing the structure of the data, predictive features in hierarchical models can be assigned to their proper levels. Such models lend themselves to hierarchical Bayes solution methods that ``share'' results of inference between groups by generalizing over the case of individual models for each group versus one model that aggregates all groups into one. 
  
    In this paper we show our work-in-progress applying a hierarchical Bayesian model to forecast purchases throughout the day at store franchises, with groupings over locations and days of the week. We demonstrate using the \textsf{stan} package on individual sales transaction data collected over the course of a year.  We show how this solves the dilemma of having limited data and hence modest accuracy for each day and location, while being able to scale to a large number of locations with improved accuracy. 
\end{abstract}

\section{Introduction}\label{sec:intro}

The field of hierarchical statistical modeling blossomed once simulation methods such as Markov Chain Monte Carlo (MCMC) made it feasible to solve models without analytic solutions (\cite{BDA3}). Approximation methods for multi-level models existed in statistics, but subsequently the field adopted these simulation methods and their Bayesian versions  (\cite{gelman2007}). Similarly in AI these techniques appeared in the form of Probabilistic Graphical Models (PGMs) (\cite{koller2009}). These graphical DAGs have spawned a bewildering variety of synonymous terms, going back to the original ``Bayesian Belief Networks,'' (\cite{Pearl2009} with current interest in causal reasoning introducing yet another term: ``Causal Graphical Models'' Concurrently hierarchical modelling is well represented in the Economics literature (\cite{geffner2022}). The similarity in data structures leads to some curiosity of cross fertilization with Statistical Relational Learning (\cite{getoor2007}).

It is our belief that these methods are widely applicable and deserve wider recognition. Most data do not come in flat tables, but come in relational tables representing different levels of aggregation. The goal of this paper is to demonstrate the value of modeling hierarchy explicitly in an actual application.

\subsection{``Point of Sale'' Demand Forecasting}

In this paper we present a sales demand model that applies over the fleet of fast-food store locations, to be used to inform real-time customer demand. Fast-food items have a short shelf-life; There is no carry-over from one prediction period to the next. Each location has control of how many products it prepares in anticipation of customer demand. Demand is uncertain, and each store needs to anticipate how many items to set aside  in anticipation of the demand in the current period. There is a trade-off: If an item is not sold punctually--- within less than an hour---it is discarded. When too many are made, at the end of the period items are discarded.  When not enough are made, customers’ demand is not satisfied, creating an “opportunity cost” of lost revenue. 

\subsection{The Current Cloud-IT infrastructure}

With the current IT infrastructure, demand is calculated centrally by a separate model for each franchise location that generates an estimate of its demand for the next period. Estimates are transmitted to the ``Point of Sales'' (POS) terminals at each location by means of the distributed ``Cloud'' application  that manages all data interaction between individual store locations and the firm.

The demand model makes up a small part of this recently implemented centralized system. After the system went into operation and a few months of data had been collected, an initial demand prediction model was built using proprietary automated time-series software. The predictions either tended toward the median value of zero, or failed to converge and our data science group was brought in to formulate a replacement model. A primary design concern is the computational cost of scaling. The intent is to deploy forecasts to thousands of stores in near-real-time. Our approach is presented here, where in Section\ref{sec:problem} we detail the problem and the available data; then in Section\ref{sec:model} explain the hierarchical model we developed. Section\ref{sec:evaluation} presents our results, where we consider these in the context of the original problem and future approaches in Section\ref{sec:conclusion}.  

\section{Description of the Problem}\label{sec:problem}

Forecasting the demand distribution—--the rate that customers arrive and the quantity of products they request promises not only to reduce cost due to minimizing waste, but also to increase revenue from anticipating times of unmet demand.
Optimizing production to meet uncertain demand is characteristic of many production and inventory problems. When as in this case, there is no inventory hold-over from one period to the next, this specifically is known as the “news-vendor” problem, (\cite{porteus2002}) a reference to a newsboy or girl who must decide how many papers to buy for the day based on their estimate of how many they will sell. 

\begin{figure}[htp]
    \centering
            \includegraphics[width=0.5\textwidth]{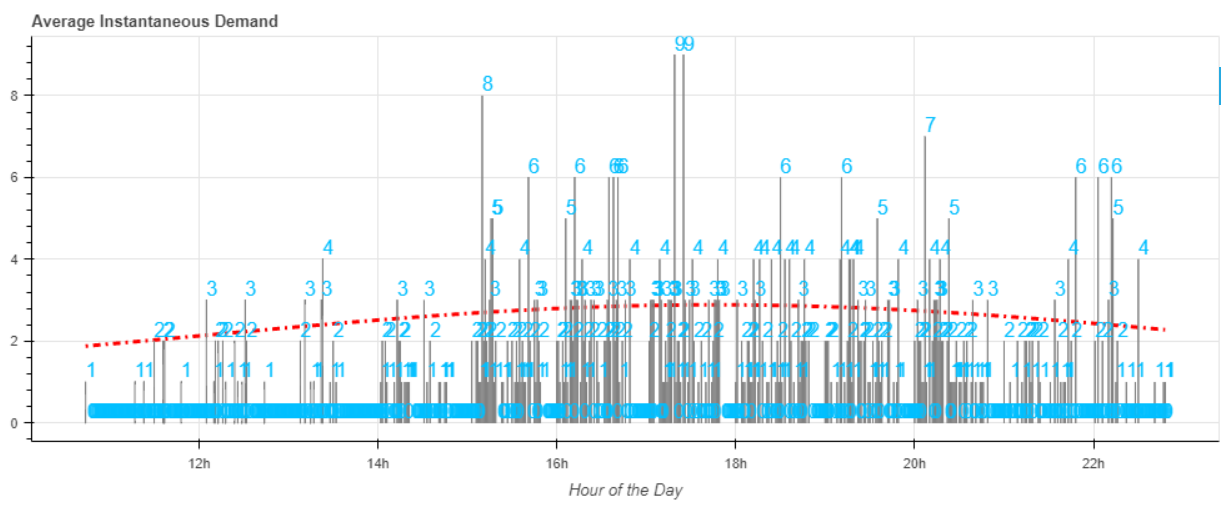}
    \caption{Sales events as time and item count at one location, for one day. The red curve is a $2^{nd}$-order fit to the log of the item count.}\label{fig:daily}
\end{figure}

Sales transactions data consists of a sequence of events in continuous time---a \emph{stochastic renewal process.} Each event consists of a random number of items ordered at a random time. The events appear as a dis-continuous function of time. A sense of the arrival-events data and how the arrival rate varies over the course of a day is shown in Figure~\ref{fig:daily}. Such event processes do not constitute a time-series in the conventional sense, confounding conventional time-series forecasting methods that inevitably fail when applied to such data.  The best one can hope for is to approximate the underlying arrival rate as a smoothly varying function as the time-series of interest.

Given the uncertainty in the arrival process even with a known, fixed arrival rate, there is a finite limit in the savings possible by any continuous function approximation to arrival events. In the simplest case one can imagine arrivals described by a Poisson distribution whose variance equals the arrival rate. In actuality arrivals tend to be ``over-dispersed'', making variance even larger. Large variance implies that any prediction based on the arrival rate is likely to be far from the actual value. For the same reason conventional time-series error estimates, such as ``MAE'' are of the order of the arrival rate parameter, and are of little use for evaluation. 

To emphasize the point, as the prediction interval for perishable inventory gets smaller, the inherent increase in variability in arrivals leads to unavoidable losses. To illustrate, if the rate is such that one customer is expected to arrive in a period there are roughly equal probabilities of no customers arriving, the one customer actually arriving, or more than one customer arriving. Thus at any instant, even with the best prediction, inventory will more likely than not be either under or over the instantaneous predicted amount.  

So given the intrinsic uncertainty of a stochastic renewal process that best describes demand for items, the best one can hope for is to predict an instantaneous demand rate.  We make this concession, apply a rudimentary regression model at the daily forecast level and focus on modelling the variability at the level of store locations and days; hence the attractiveness of a hierarchical approach. The hierarchical model is more interpretable, and since the upper level data is more concise by a couple orders of magnitude, the model can scale to thousands of stores. 

Short of being able to model the larger in-store production optimization problem, we leave it up to the store manager to decide on the actual production policy as informed by our prediction of the instantaneous rate of demand.

\begin{figure}[htp]
    \centering
            \includegraphics[width=0.5\textwidth]{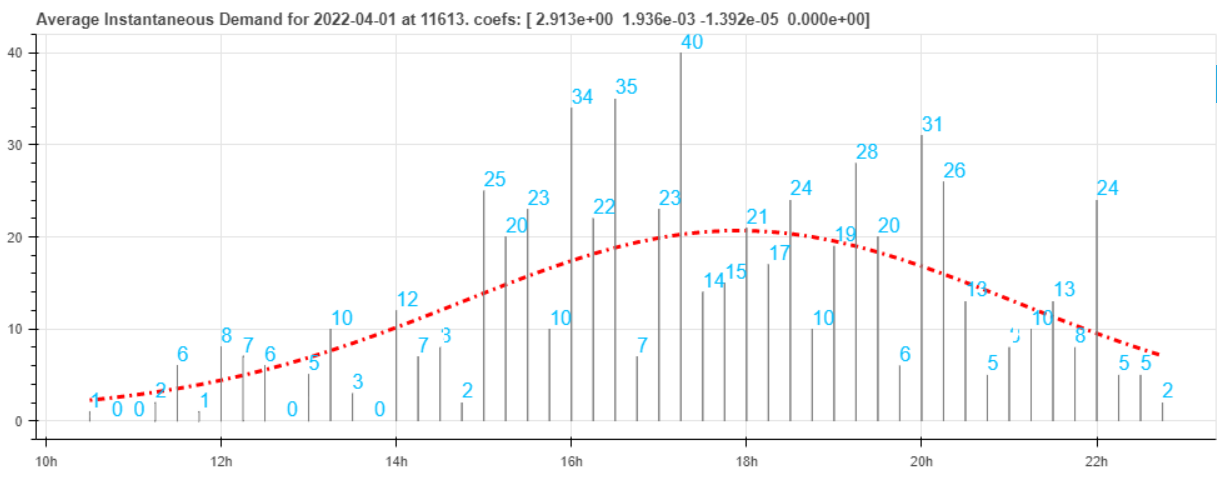}
    \caption{Binned sales events over 15 minute intervals for the same one location and day. The red curve is a $2^{nd}$-order fit to the log of the binned item count. }\label{fig:binned_daily}
\end{figure}

\subsection{Forming the Hierarchy}

At the lowest level of the hierarchy are the individual sales transaction events consisting of 1,271,639 transactions at 49 stores over the course of about half a year, for which we have data from 243 days of data, about 150 which are usable. We group the daily samples by \textsf{location} and \textsf{day\_of\_week}, and fit a parameterized arrival rate curve to that location-day to create the upper level data. At the upper level of the hierarchy, the number of samples is reduced significantly, to $|locations| \times |days| \times |local\_model\_coefficients|$. There are 8649 ``location-day'' records, one for each day and location, each described by 3 coefficients, a reduction to 2\% of the original data size! The schema for the upper level data is show in Table\ref{tab:hierarchy}. The relations among the groupings---location, day-of-week and transaction are analogous to those in a relational data schema, with \textsf{location} and \textsf{day\_of\_week} as common keys. 

We form two upper-level groupings; by location and by day-of-week. Each grouping could have co-variates; for instance, \textsf{location} could be influenced by regional economic conditions, and \textsf{day\_of\_week} by holidays. These extensions to the model fit naturally into our formulation and arguably could substantially improve predictions, but here we focus on the basic hierarchical design, and inclusion of upper-level co-variates is not considered. 

\subsection{Data Sources and Preparation}

We were provided by the client with a snapshot of the aforementioned 1,271,639 POS daily sales transaction records over the past year. Each sales transaction event records the time and number of items purchased by a customer with one-minute resolution. The sales record fields are shown in Table~\ref{tab:fields}.

This transaction event data is binned on 15-minute intervals to generate counts on a fixed time grid reducing the transactions to 357,224 records.  Intervals when no sales occur are preserved and recorded as zero counts.  Binning is necessary to fit an estimate of the instantaneous transaction rate not biased by the timing of transaction events. 

To create the upper hierarchy levels the binned data for each day and location were summarized by a curve fit by 3 coefficients. Figure~\ref{fig:binned_daily} is an example of a curve for one case. These cases make up the coefficients data set used for the hierarchical model, with fields described in Table~\ref{tab:hierarchy}. We assign records in the coefficients data set randomly with probability $1/2$ to train and test sets.

Inspection of the data over the year's interval showed a slight increasing demand trend plus a substantial interval of missing data for the middle quarter of the year. Our initial analysis, using the early data for train and later for test revealed that the data set was not stationary. This approach confounded any test results on the data split and was abandoned.

\begin{figure}[htp]
    \centering
            \includegraphics[width=0.5\textwidth]{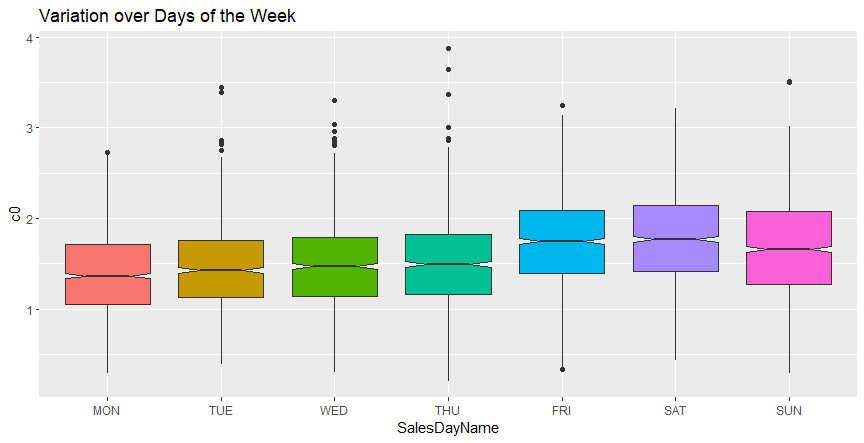}
    \caption{These barplots show the variation in the $c0$ coefficient for each day of the week. The values for each day varies widely, with some weekly variation peaking on Saturdays.}\label{fig:var_days}
\end{figure}

\begin{figure}[htp]
    \centering
            \includegraphics[width=0.5\textwidth]{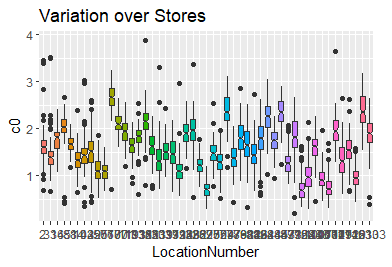}
    \caption{Sales events as time and item count at one location, for one day. The red curve is a $2^{nd}$-order fit to the log of the item count.}\label{fig:var_stores}
\end{figure}

Characteristic of hierarchical models, our premise is that the variation within the two groups, \textsf{location} and \textsf{day\_of\_week} explain the variation in the coefficients data. This boxplot, Figure\ref{fig:var_days}, shows the variation of the first coefficient $c0$, between and within days of the week. A similar plot for stores is shown in Figure\ref{fig:var_stores}. As one might expect, the finer-grained store grouping shows less within-group variation and more between-group variation than among days of the week. 

\begin{table}[ht]
\centering
\caption{Fields in the sales transaction dataset.}
\label{tab:fields}
\resizebox{0.9\columnwidth}{!}{
\begin{tabular}{|c|c|c|c|}\hline
     \textbf{Field name}&  \textbf{type}& \textbf{Description}\\\hline\hline
     \textbf{LocationNumber}& int& Vendor identifier\\\hline  
     \textbf{SalesDayName}&  string& Day of the week\\\hline
     \textbf{DailyMinutesOpen}& int& Daily operating minutes\\\hline
     \textbf{DateTimePlaced}& datetime& Transaction timestamp\\\hline  
     \textbf{SalesAsMinutes}&  double& Minutes since opening\\\hline
     \textbf{Quantity}& int& Number of items purchased\\\hline
\end{tabular}}
\end{table}


\begin{table}[ht]
\centering
\caption{Fields in the hierarchy dataset.}
\label{tab:hierarchy}
\resizebox{0.9\columnwidth}{!}{
\begin{tabular}{|c|c|c|c|}\hline
     \textbf{Field name}&  \textbf{type}& \textbf{Description}\\\hline\hline
     \textbf{LocationNumber}& int& Store identifier\\\hline  
     \textbf{Day}& int& Calendar Day\\\hline  
     \textbf{SalesDayName}&  string& Day of the week\\\hline
     \textbf{Coefficient0}& double& Constant\\\hline
     \textbf{Coefficient1}& double& Slope\\\hline  
     \textbf{Coefficient2}&  double& Curvature\\\hline
\end{tabular}}
\end{table}



\section{Sales Demand Model}\label{sec:model}

Given the stochastic nature of minute-to-minute demand, most of the value in prediction comes from how this rate depends on ``upper level'' factors. As mentioned at the upper levels the data is grouped by location and day of the week. Our stochastic renewal process assumes  that variation from minute to minute is not predictable from preceding events. We approximate the instantaneous demand rate function for each day, and model it by three parameters:

\begin{itemize}
    \item [$c0$]: the average daily demand,
    \item [$c1$]: the daily trend,
    \item [$c2$]: the ``peakedness'' of the rate.
\end{itemize}

We arrive at these values by regressing the centered daily sales arrivals using a $2^{nd}$ order polynomial---equivalent to an orthonormal basis of the first 3 Lagrange polynomials---one has 3 independent values that describe each day and store, of which there are, as mentioned, 8649 instances. $c0$ coefficients tend to small positive values, as Figures\ref{fig:var_days} and \ref{fig:var_stores} show. $c1$ and $c2$ coefficients are several orders of magnitude smaller, clustering around zero. This suggests that the daily transaction rate is relatively constant with minor tendencies to either rise or fall during the day.  Each regression might encompass a few hundred points, but some instances are sparse and include just a few points.  The ability for hierarchical Bayesian methods to ``share'' information among units from upper levels improves predictions in these cases. 

The regression step is computationally inexpensive and, in a production system, would be done at each location, offloading much of the central computing load to make it possible to scale the model to 1000s of locations.  Certainly more sophisticated local models can be applied, but we use this approximation that we pre-calculate of the lowest level of the hierarchy to focus attention on the core problem. 

\subsection{The Graphical Model}\label{sec:cgm}

\begin{figure}[htp]
    \centering
            \includegraphics[width=0.54\textwidth]{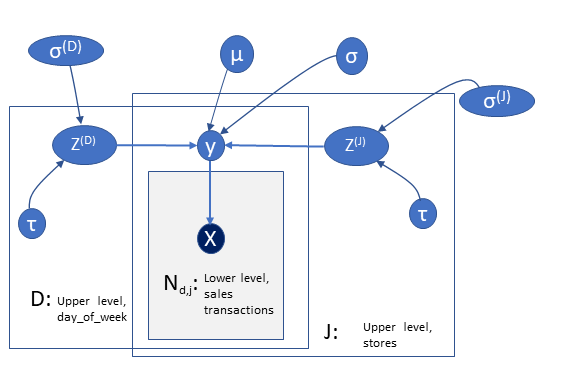}
    \caption{The graphical model network for a model with two upper levels, one for $J$ locations, the other for $D$ days of the week. Each node represents a distribution in the generative model. Nodes on a plate are replicated by the plate index. Overlapping plates indicate replication by the product of the indexes. Subscripts for nodes on a plate are implied. Nodes outside any plate are global parameters. Sales transaction data $x$ is modelled separately as 3 component parameters, implying 3 separate models for $y$.  }\label{fig:net}
\end{figure}

The network diagram, Figure~\ref{fig:net} shows the graphical model using plate notation. Each upper level is represented by a separate overlapping plate. The shaded inner plate is solved separately by the regression step and fed into the upper level model.  This approximation can be removed in the future, but it is adequate to demonstrate the benefits of the hierarchy.  In the statistical literature this design is called a ``two-way random effects model.''\footnote{See \cite{gelman2007}, p.245. Despite using it, the author argues against use of the the term ``random effects'', since the Bayesian formulation has no need for the random effects -- fixed effects distinction. }


The model equation, with the distribution assumptions is described by these four equations. The superscripts, $D$ and $J$ indicate the day and location grouping levels.  Any parameter without an explicit prior (e.g. $\mu$) is assumed to have a uniform prior. 
\begin{align}
y_{dj} &= z^{(D)}_d + z^{(J)}_j + \mu\\
z^{(D)}_d &\sim \mathcal{N}(\tau_d, \sigma^{(D)})\\
z^{(J)}_j &\sim \mathcal{N}(\tau_j, \sigma^{(J)})\\
y &\sim \mathcal{N}(y_{dj}, \sigma)
\end{align}

We run this model separately for each class of coefficients, 
 $(c0_{dj}, c1_{dj}, c2_{dj})$ by substituting them for the $y$ variable, to create three versions of the model. 

The inputs to the model are the $y$, indexed by location $j$ and day-of-week $d$. The model estimates the distribution of the contribution of each day-of-week $z^{(D)}$, and location $z^{(J)}$. Four global parameter distributions are output for $\sigma^{(D)}$,  $\sigma^{(J)}$, $\sigma$, and $\mu$. The MCMC simulation of the model outputs marginal posteriors over each of the $4 + 7 + 49$ parameters conditioned on the observed data $y$.

The \textsf{stan} file for these equations in shown in Figure~\ref{fig:stanfile} in the Appendix. 

\subsection{Training}\label{sec:training}

The \textsf{stan} code to run the model is shown in Figure~\ref{fig:stan}, showing the sampling data read as inputs, and the parameters (\textsf{par}) as outputs.  An MCMC training run on the train dataset of 4302 samples completes in less than 1 minute per chain on a current laptop. We observed less-than-linear increases in processing time with larger samples, when running on the combined train and test dataset. 

Runtime for each model was less than 4 minutes. 
To assure convergence for $c1$ and $c2$ values the $y$ data was rescaled to have mean near 1. 

\begin{figure*}
\begin{lstlisting}[language=R]
model <- stan_model("regression_coef.stan")
rc_fit <- sampling(model, 
    data=c("N","D", "J", "d_id", "j_id", "y"), iter=4000, chains=4)
params <- summary(rc_fit, 
    pars=c("mu", "s_d", "s_j", "s_elipson", "z_d", "z_j"))$summary
\end{lstlisting}
\caption{\textsf{rstan} code to run the model.}\label{fig:stan}
\end{figure*}

\textsf{Stan} returns several diagnostics including tests for sampling convergence, for which our model has no issues. Among other diagnostics is a log-likelihood measure useful for comparisons among runs: 
\begin{verbatim}
# mean   se_mean       sd   
# lp__   1410.121     0.1888
\end{verbatim}

Another sanity check is to see the estimates of variance, as one would in conventional analysis of variance.  We see that the sum of variances closely approximates the $y$ variance.

\begin{align*}
    0.445 &= \sqrt{\sigma^{{(D)}^2} + \sigma^{{(J)}^2} + \sigma^2}\\
    &\mbox{compared to}\\
    0.419 &= \sigma_y\\
\end{align*}

The fraction of explained variation is 
\begin{equation*}
     R^2 = 1 - \sigma^2 / \sigma_y^2 = 0.638
\end{equation*}

\section{Evaluation}\label{sec:evaluation}

Despite the MCMC results returning the full sample marginals for all estimated parameters, we use a conventional accuracy measure for prediction of the coefficients. 
\emph{Bias} is simply the sum of the difference in the means of predicted to actual. 
\emph{RMSE} is the square root of the average squared deviation of the differences between predicted and actual. The model predicts the three coefficient values for each of the upper-level units: the 49 locations and 7 days of the week. These can be compared to a baseline prediction, of just using the average coefficient value for the hold-out set over those units. Those averages are exactly the means of the boxplots shown in Figures\ref{fig:var_stores} and \ref{fig:var_days}  This is a challenging test; the average hold-out values are themselves an accurate predictor. 

Ideally one would want to evaluate improvements at the transactions level and not only at the level of the coefficients fit to the daily transactions.  As argued, the variation of events around the instantaneous demand rate is so large it would obscure any improvements to the coefficients that determine the instantaneous demand rate. 

First as a sanity check we compute \emph{bias} and \emph{RMSE} error for ``test on train'', for $c0$, by predicting from the values that were trained on.  These errors are negligible, as expected: 

\begin{verbatim}
Test on 'train' 
   $bias
   -0.0003

   $rmse
   0.00876
\end{verbatim}

Full evaluation tests were run for the hierarchical model, for both the location predictions and day-of-week predictions for the 3 separate components, and compared to baseline averages.  The smaller of the errors between the baseline average prediction and the hierarchical prediction is shown in bold, in Table\ref{tab:hierarchy}.

\begin{table}[ht]
\centering
\caption{RMSE Errors.}
\label{tab:hierarchy}
\resizebox{0.9\columnwidth}{!}{
\begin{tabular}{|c|c|c|c|}\hline
     \textbf{Coefficient}&  \textbf{Group}& \textbf{Average}& \textbf{Hierarchy}\\\hline\hline
     c0& Location& 0.0398& \textbf{0.0381}\\\hline 
     c0& Day-Of-Week& \textbf{0.0277}& 0.252\\\hline 
     c1& Location& 2.860e-4&\textbf{2.838e-4}\\\hline 
     c1& Day-Of-Week& 6.61e-5& \textbf{6.46e-5}\\\hline
     c2& Location& 1.012e-06&\textbf{0.980e-6}\\\hline 
     c2& Day-Of-Week& 2.858e-7& \textbf{2.614e-7}\\\hline
\end{tabular}}
\end{table}

Except for the Day-of-week prediction for $c0$ coefficients the hierarchical model incrementally outperforms the baseline. Observing that using the average value as a predictor obtains an RMSE error of just a few percent, this is an impressive improvement by the hierarchical simulation model and recommends its use in production.  

\begin{figure}[htp]
    \centering
        \centering
            \includegraphics[width=0.4\textwidth, height=0.4\textwidth]{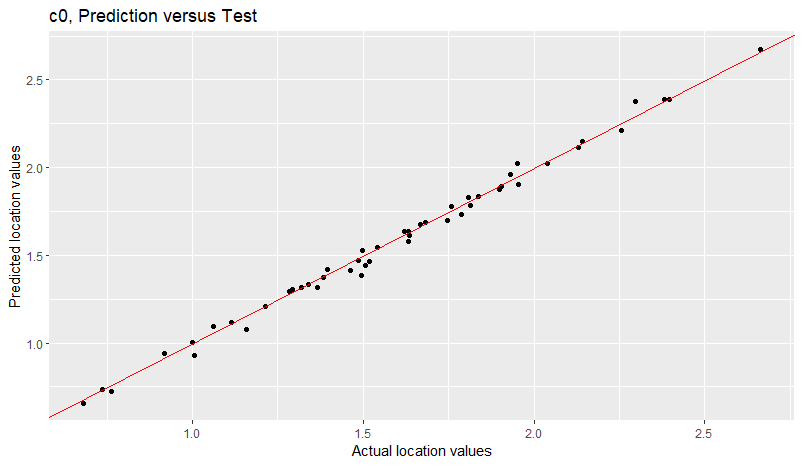}
    \caption{Predicted \textit{coefficient0} versus location means from the actual test set as a predictor. Each point is one location. The red line shows $x=y$.}\label{fig:predicted0}
\end{figure}

\begin{figure}[htp]
    \centering
        \centering
            \includegraphics[width=0.4\textwidth, height=0.4\textwidth]{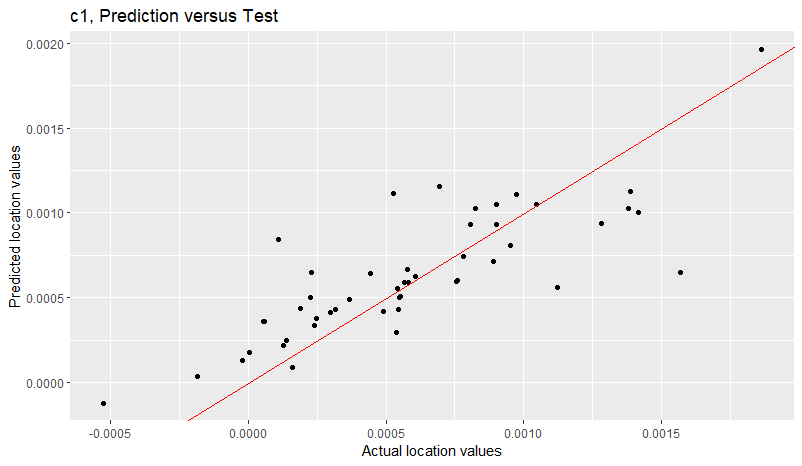}
    \caption{Predicted \textit{coefficient1} versus location means from the actual test set.}\label{fig:predicted1}
\end{figure}

\section{Discussion}

\subsection{Alternative approaches}

Popular today are numerous deep-learning based time-series forecasting tools such as\cite{Kuznetsov2018}, which considers the whether multiple series should be learnt in a common model or separately. Although this approach may approximate the ability to generalize over multiple (e.g. the 324 time-of-day location) series, it has none of the advantages of the method presented. We believe that such a method invests vast effort in learning from an stochastic event series, where simple smoothing methods suffice, at the expense of analyzing the effects of groupings and the contextual variables on the more parsimonious upper-level data where interesting effects occur. It would scale to the thousands of locations at only comparatively large computational cost. 

An ``off-the-shelf'' option for hierarchical modelling is to use a current spatial-temporal forecasting tool instead, such as FOST. FOST combines traditional and ML-based temporal modeling with spatial modeling via graph neural networks.
FOST requires two data inputs: time series data and “spatial” data in the form of a list of directed graph edges and their weights. Edges can represent truly spatial data, such as distances or travel times. Additionally, edges can represent relationships such as correlations between the items represented by graph nodes. If the nodes represent products for sale, then edges can represent cannibalization or halo effects. A halo effect is when reducing the price of one product increases sales of a related product---a cannibalization effect is the opposite.

As for spatial effects, it depends on how close together stores are. Thus we could create a graph of locations, with edges weighted by distance or travel time. Alternatively, we could create a graph with edges weighted by the correlation of sales between pairs of stores. 

Effects among locations expressed on a graph structure is a level of model complexity not captured by a hierarchy. Relations in a hierarchy are mediated by presence of common parents, not by pair-wise relations. The choice of model depends speculatively on the relevance of relations that would be expressed in a graph, the feasibility of predicting from event process data, and scalability to the number locations.

\subsection{Discussion}

The apparent non-stationarity of the current dataset has been managed by ignoring the temporal separation of test and train splits. Arguably this is of minor consequence because the groupings by day-of-week and location averaged over the year  do not relate to each other temporally. 

We have many ideas how to extend the current model that at this point just demonstrates the feasibility of this approach. Extensions of the Graphical Model could bring in temporal predictors to see if they do have any effect.  One could test if the conditional independence of day-of-week and location is valid, but a priori there is little reason to think it is not. Most interestingly one could introduce contextual variables to the upper levels for economic, demographic, weather, holiday, spatial and other effects that are presumed determinants of demand. 

Simulation methods such as MCMC do not always converge in fixed time, so a more reliable inference approximation method would be preferable.  Using Expectation Propagation in place of MCMC, as for instance implemented by \emph{infer.net}\cite{Wang2011} is worth trying. 


By only forecasting recorded transactions we overlook demand by customers turned away because no items were available for sale. These events are not captured. This biases estimates downward. An outstanding challenge in forecasting is to predict unmet demand, so one would know that a sale would be missed because items were not available. One possible way to address this is to model the censoring process. We assume a distribution for transactions, say Poisson, or a more general renewal distribution, then ``extrapolate'' the distribution to estimate extreme values not seen in the data. 

Alternatively if one had sales transactions and production amounts data then the the full demand distribution could  be estimated. Then one could tell when production limits sales (e.g. when a store is sold-out). So by analyzing sales data when stores are not sold out the full demand curve could be fit to estimate unmet sales and remove the source of forecast bias.

An obvious oversight in this model formation is that lack of 
a ``supply'' model to complement the ``demand'' model. A comprehensive model would recommend the optimal production schedule for perishable items. For that we'd need a corresponding  production model to complement this demand model. The value of improvements due to the model cannot be evaluated without also observing and modeling real-time production in which case one would also need historical data on actual production.

\subsection{Conclusion}\label{sec:conclusion}

We've demonstrated promising results on an initial framework for a hierarchical model of sales transactions over groupings of locations and days. 
As in many applied modelling problems, there are natural groupings of the samples that can be expressed hierarchically, avoiding the need to pre-process the data to de-normalize it such as done by ``one-hot'' encoding. Such problems translate directly into Bayesian  Graphical Models for which there are now mature solution tools. The advantages of formulating hierarchical as Graphical Models are several. 1. Groups with limited data benefit from inference from similar neighboring groups, increasing the effective sample size of groups. The degree of interaction is a consequence of the sample size within groups. 2. Co-variates can be applied directly to the relevant upper levels. 3. Models may be federated at lower levels for distributed processing. 4. The full value of Bayesian methods in terms of explainability and the derivation of true predictive distributions can be exploited.


\begin{acknowledgements} 
    Without the express help of my data science colleagues, and numerous sales, marketing, and engineering staff both in-house and at the client, none of this would be possible. 

\end{acknowledgements}

\bibliography{uai2022-bib}

\appendix

\begin{figure*}
\section*{Appendix}
\begin{lstlisting}
// This Stan program defines a simple model, with a two separate
// groups 
// -- J locations
// -- D days of the week. 

data {
  int<lower=0> N;                 // row count
  int<lower=0> D;                 // The number of days of the week 
  int<lower=0> J;                 // The number of locations
  int<lower=1, upper=D> d_id[N];  // map rows into days of week
  int<lower=1, upper=J> j_id[N];  // map rows into locations. 
  vector[N] y;    //  outcome -- the predicted value for that dow & location. 
}

// The parameters accepted by the model. Our model
// includes a eta and sigma for each effect
parameters {
  real mu;
  vector[D] eta_d;
  vector[J] eta_j;
  real<lower=0> s_d;
  real<lower=0> s_j;
  real<lower=0> s_elipson;
}
transformed parameters{
  vector[J] z_j;
  vector[D] z_d;
  vector[N] yhat;
  
  z_j = s_j * eta_j;  // random effects
  z_d = s_d * eta_d;
  
  for (i in 1:N) {
    yhat[i] <- mu + z_d[d_id[i]] + z_j[j_id[i]];
  }
}

// The model to be estimated.
model {
  eta_d ~ normal(0,1);
  eta_j ~ normal(0,1);
  y ~ normal(yhat, s_elipson);
}    
\end{lstlisting}
\caption{File \textsf{regression\_coef.stan} of a two-way random effects model}\label{fig:stanfile}
\end{figure*}

\end{document}